\documentclass[epj]{svjour}
\usepackage{graphics}
\begin{document}
\title{Bimodal pattern in the fragmentation of Au quasi-projectiles }
\author{M. Bruno\inst{1},  F. Gulminelli\inst{2}\thanks{\emph{Member of the Institut Universitaire de France}}
, F. Cannata\inst{1}, M. D'Agostino\inst{1}, F. Gramegna\inst{3}, G.
Vannini\inst{1}
}                     % Do not remove
\institute{Dipartimento di Fisica dell'Universit\`{a} and INFN,
Bologna, Italy \and LPC Caen (IN2P3-CNRS/EnsiCaen et
Universit\`{e}), Caen, France \and INFN, Laboratori Nazionali di
Legnaro, Italy}
\date{Received: date / Revised version: date}
\abstract{ Signals of bimodality have been investigated in
experimental data of quasi-projectile decay produced in Au+Au
collisions at 35 AMeV. This same data set was already shown to
present several signals characteristic of a first order,
liquid-gas-like phase transition. For the present analysis, events
are sorted in bins of transverse energy of light charged particles
emitted by the quasi-target source. A sudden change in the
fragmentation pattern is observed from the distributions of the
asymmetry of the two largest fragments, and the charge of the
largest fragment. This latter distribution shows a bimodal behavior.
The interpretation of this signal is discussed.
 % end of PACS codes
} %end of abstract
\titlerunning{Bimodal pattern in the fragmentation of Au quasi-projectiles}
\authorrunning{M. Bruno et al.}
\maketitle
\section{Introduction}
\label{intro}
The existence of different phases for infinite nuclear
matter is predicted by theoretical calculations since the early 80's~\cite{history}.
Then, the possibility of observing a nuclear liquid-gas phase transition in the
laboratory has been deduced from several experimental observations
associated to the multi-fragmentation of finite nuclei.
These observations indicate the occurrence
of a state change in finite nuclei, which is interpreted to be
the finite system counterpart of a phase transition~\cite{chomaz}.

Many of these signals are qualitative, and therefore can\-not give
information on the detailed tra\-jec\-to\-ry (in terms of pressure,
volume, temperature, isospin) followed by the system from one phase
to the other. Other signals give in principle quantitative
information, but can be distorted. Indeed the products of the
reactions are detected asymptotically and not at the production
time, and therefore they need to be corrected for secondary
decay~\cite{decay}. These corrections are, at least partially, model
dependent and induce systematic errors which are difficult to
estimate quantitatively.

To overcome these difficulties, it is important to perform a
systematic study of different phase transition signals. The best
would be to exploit new generation 4$\pi$ apparatuses, in order to
be able to investigate several signals at the same time, with the
same experimental data samples, and with a complete or
quasi-complete detection~\cite{praga}. Waiting for these new
apparatuses, some of the signals indicating a phase transition have
been obtained with measurements performed by the Multics~\cite{mcs},
together with the Miniball~\cite{mini}, multi-detectors. In the last
few years we have investigated in detail the properties of
quasi-projectiles detected in Au + Au reactions at 35 A.MeV, with a
fixed source charge, and at different excitation energies. The
following signals have been obtained:
\begin {enumerate}
\item the average size of the heaviest fragment (tentatively associated to the
liquid part) decreases for increasing excitation energy of the
nuclear system~\cite{dag} with a power law distribution of exponent
$\beta\approx 0.31$;
\item temperature measurements result compatible~\cite{mil,reliability} with a "plateau" in
the caloric curve~\cite{caloric}; \item critical exponents have been
extracted~\cite{dag}, close to the values expected within the
liquid-gas universality class;
\item the size distribution presents a scaling
$\grave{a}$ la Fisher~\cite{critical};
\item interaction energy fluctuations, corrected for si\-de-fee\-ding,
were shown to overcome the statistical expectation in the canonical
ensemble, corresponding to a negative branch of the microcanonical
heat capacity for a system in thermodynamical
equilibrium~\cite{negative}.
\end {enumerate}

Some of these signals are coherent with the findings of other
experimental collaborations with different data sets
\cite{ma,natowitz,eos,isis}. In particular, the last two signals
have been confirmed in central reaction measurements performed with
the Multics~\cite{dag2,bruno} and with the
Indra~\cite{reliability,leneindre} apparatuses. Some of these
behaviors were also observed in other finite physical systems
undergoing a transformation interpreted as a first order phase
transition, namely in the melting of atomic
clusters~\cite{negclu,turchi} and in the fragmentation of hydrogen
clusters~\cite{neg3}.

Recently~\cite{PRE}, a new topological observable has been proposed
to recognize first order phase transitions. When a finite system
undergoes such a transition, the most probable value of the order
parameter changes discontinuously, while the associated distribution
is bimodal close to the transition point, i.e. it shows two separate
peaks, corresponding to the two different
phases~\cite{binder,dasgupta}. In the case of transitions with a
finite latent heat, this behavior is in agreement with the Yang-Lee
theorem for the distribution of zeroes of the canonical partition
sum in the complex temperature plane~\cite{zeroes}, and equivalent
to the presence of a curvature anomaly in the microcanonical entropy
$S(E)$~\cite{intruder}.

Since many different correlated observables can serve as order
parameters in a finite system, the task is to choose an order
parameter which can be accessible from the experimental
side~\cite{WCI}. This is the case for observables related to the
measured charges. The INDRA collaboration~\cite{indra} has proposed
as order parameter the
 variable Z$_{sym}=\frac {Z_1-Z_2} {Z_1+Z_2}$, where
Z$_1$ and Z$_2$ are the charge of the largest and the second largest
fragments detected, in each event, in the decay of an excited
source. An indication of a bimodal distribution was obtained for
this quantity. Signals of bimodality in different observables have
been obtained in experiments with different projectile-target
combinations, and in different energy ran\-ges~\cite{ma,WCI,bell}.

In ref.~\cite{WCI} it has been pointed out that the variable
Z$_{sym}$ can present a spurious bimodality in small
three-dimensional percolation lattices close to the percolation
threshold. This behavior is due to finite size, and makes a
bimodality in Z$_{sym}$ an ambiguous signature of the transition. On
the other side, the size $A_1$ or charge $Z_1$ of the largest
fragment have distributions which for any lattice size are
consistent with the critical percolation behavior~\cite{campi}.
These observables were then suggested as more apt to discriminate
between a first order phase transition, a critical phenomenon, and a
smooth cross-over.

In this paper we investigate if these different signals are present
in our data. We also discuss the relation of the bimodality signal
with other phase transitions indicators obtained for our data.

\section{The experiment}

\label{sec:1} The measurements and the analysis have been
extensively described elsewhere~\cite{dag}. Here we recall that the
measurements were performed at the K1200-NSCL Cyclotron of the
Michigan State University. The Multics~\cite{mcs} and
Miniball~\cite{mini} arrays were coupled to measure light charged
particles and fragments with a geometrical acceptance of the order
of 87\% of 4$\pi$. The events have been recorded if at least two
different modules have been fired. Similarly to Ref.~\cite{indra},
the selection of the quasi-projectile (QP) source have been made by
a shape analysis, keeping all the fragments with $Z\geq3$. The
fragments have been considered as belonging to the QP, if forward
emitted in the ellipsoid reference frame. The complete source has
been obtained by doubling the forward emitted light particles in the
backward direction, in order to minimize the contamination of light
particles emitted by a possible mid-velocity source. At the end of
this procedure, only events with total charge within 10\% of the Au
charge have been considered for the analysis, in order to study the
decay of a well detected constant size source, in a wide range of
excitation energies.

\begin{figure}[htbp]
\begin{center} \resizebox{.8\columnwidth}{!}{\includegraphics{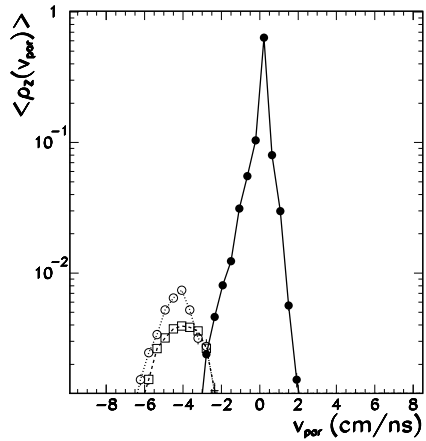}}
\caption{Charge density distribution of QP Au+Au events, as a function of the fragment velocity, along an axis parallel to the QP velocity.
The continuous line (full
points) represents the charge density for fragments accepted for the
QP, the dashed line (open squares) for fragments of the QT, the
dotted line (open circles) is a filtered simulation of a QT source
symmetric to the experimentally detected QP.} \label{vpar}\end{center}
\end{figure}

In order to visualize the source characteristics in the selected
events, the fragment (Z$\geq 3$) charge density
distribution~\cite{lecolley} is shown in Fig.~\ref{vpar} as a
function of the fragment velocity in the QP reference frame. The
ensemble averaged charge density $\langle\rho_Z(v_{par})\rangle$ is
defined as

$$ \langle\rho_Z(v_{par})\rangle = \Big\langle {\sum Z(v_{par}) \over
\sum Z} \Big\rangle $$

where $\rho_Z(v_{par})$ is the event-by-event distribution in the
velocity $v_{par}$ for the collected charge fraction. This
observable represents the distribution of the collected charge bound
in fragments along the direction of the QP velocity. In
Fig.~\ref{vpar} the continuous line (full points) represents the
charge density for fragments accepted for the QP, the dashed line
(open squares) for fragments of the QT. This latter is consistent
with the filtered simulation of a QT source symmetric to the QP
(dotted line - open circles). QP and QT can be easily recognized,
showing that the imposed conditions are effective to select events
where the contamination of a mid-velocity source is negligible (for
more details see Refs.~\cite{dag,reliability}).

\begin{figure}[htbp]
\begin{center} \resizebox{.99\columnwidth}{!}{\includegraphics{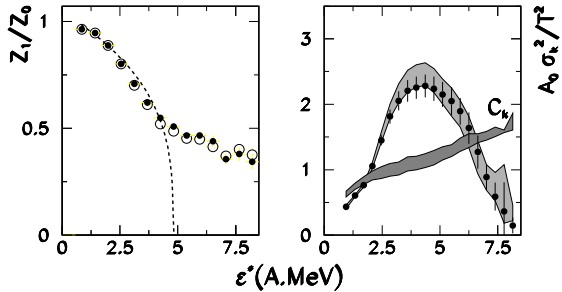}}
\caption{Left panel: charge of the largest fragment Z$_1$ normalized
to the size of the source Z$_0$ as a function of the QP calorimetric
excitation energy. The symbols refer to event selected with (open
circles) and without (full points) the constraint on the velocity of
the largest fragment. The dashed line is a power law with exponent
$\beta$ = 0.31. The right panel shows the normalized partial energy
fluctuations for QP events selected with (grey contours) and without
(full points) the constraint on the velocity of the largest
fragment. The estimation for the canonical heat capacity C$_k$ is
also shown~\protect\cite{negative}.} \label{cneg}\end{center}
\end{figure}

The characteristics of these events have been examined by analyzing
the isotropy of the fragment angular distribution  in the
quasi-projectile reference frame~\cite{negative,dag2}, and by comparing
the data to predictions of a statistical multifragmentation model~\cite{SMM}.
The general conclusion is that an important degree of
equilibration appears to be reached by the excited quasi-projectile
sources in the whole range of excitation energies.
For more details, see Refs.~\cite{dag,reliability,negative,dag2}.

In Refs.~\cite{dag,reliability} a further condition to characterize
the QP was added, i.e. that the velocity of the heaviest fragment is
larger than 75\% of the projectile velocity. This condition indeed
has been replaced with the "completeness" of the
event~\cite{WCI,indra} which reflects on the limitations on the
parallel momentum, since it results less correlated to the variables
we want to study. This does not affect the distribution shown in
Fig.~\ref{vpar}, and all the signals of phase transition do not
change significantly. To quantify this statement, we present in
Fig.~\ref{cneg} the power-law behavior of the average charge of the
largest fragment, normalized to the charge of the source, as a
function of the excitation energy, and the normalized partial-energy
fluctuations, leading to the estimate of a negative branch for the
microcanonical heat capacity~\cite{negative}. The power law in the
Z$_1$ distribution and the partial energy fluctuations are very
scarcely affected by the condition on the velocity of largest
fragment, apart from some very small variations at the higher
energies~\cite{reliability,negative}.

These observations mean that the detection apparatus appears
especially effective in the complete detection of purely binary
collisions. The same is not true for all reaction mechanisms, which
may need different detection systems to be addressed. Only a
fraction of well detected peripheral collisions can be interpreted
as the independent statistical decay of two isotropic
sources~\cite{colin}. For instance within the INDRA apparatus it has
been pointed out that for 80 A.MeV Au+Au collisions, these events
represent about the 20\% of the total number of complete
events~\cite{leneindre2}, and depend on the selection criteria
adopted~\cite{bonnet}. In our case the statistical events represent
about the 30\% of the measured events as reported in
Ref.~\cite{bo2000}, and as can be inferred from Fig.~2b) of
Ref.~\cite{dag}. The different in the percentage of statistical
events could also be due to the different trigger conditions of two
and four modules fired, used in our and Indra measurements,
respectively.

\section{Signals of bimodality}
\label{sec:2}

In the liquid-gas phase transition, the largest fragment detected in
each event is a natural order parameter, because of its correlation
with the particle density in the grancanonical
ensemble~\cite{PRE,big}. The variable proposed in
Ref.~\cite{indra,rivet} in turn is trivially correlated to the
largest cluster size, and in addition brings further information on
the global fragmentation pattern. This means that it should be
possible to observe a bimodal distribution for the charge of the
largest fragment or the asymmetry, if one considers a system close
to the transition temperature.

\begin{figure}[htbp]
\begin{center}
\resizebox{0.99\columnwidth}{!}{\includegraphics{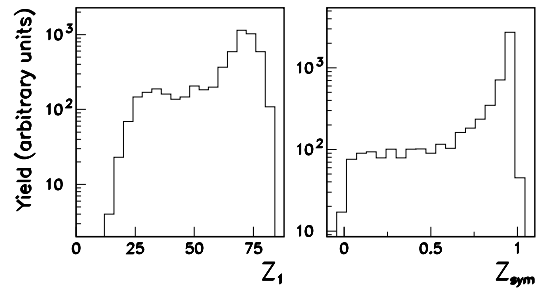}}
\caption{Distribution of the largest fragment charge Z$_1$, (left
panel) and of the asymmetry between the two largest fragments
Z$_{sym}$ (right panel) for the whole set of QP events.}
\label{whole}
\end{center}
\end{figure}

The global distributions of Z$_1$ and Z$_{sym}$ are shown in
Fig.~\ref{whole}, for all QP events selected as explained in section
2. Because of the impact parameter geometry, this distribution is
clearly dominated by peripheral collisions at low deposited energy,
leading to a heavy Z$_1\approx 75$ residue with a large asymmetry
Z$_{sym}\approx 0.9$.

We can however also notice the presence of a large bump, corresponding
to much lighter residues and much more symmetric fragmentation patterns.
For this bump to be interpreted as an indication of bimodality,
we should show that:
\begin{itemize}
\item the two different decay patterns
can be obtained in the de-excitation of the same source,
\item they correspond to the same temperature.
\end{itemize}
Let us first concentrate on the source definition. As we have
already stressed in the last section, we are considering only events
with a detected charge in the forward QP hemisphere close to the
original $Au$ charge. This guarantees a good detection, but does not
constrain the reaction mechanism or the number of sources, since the
system is symmetric. Fig.~\ref{vpar} shows that the selected events
are consistent with a purely binary kinematics, meaning that the
bump at low charge shown by Fig.\ref{whole} cannot be ascribed to a
reduced size of the excited source. However Fig.~\ref{vpar} is
obtained with the whole set of events, which is largely dominated by
peripheral collisions. We may then wonder whether a (small)
contamination of central collisions, leading to an important
stopping in the center of mass, may be responsible of a decrease of
the QP source size in the dissipative reactions corresponding to the
low Z$_1$ bump.

\begin{figure}[htbp]
\begin{center}
\resizebox{0.9\columnwidth}{!}{\includegraphics{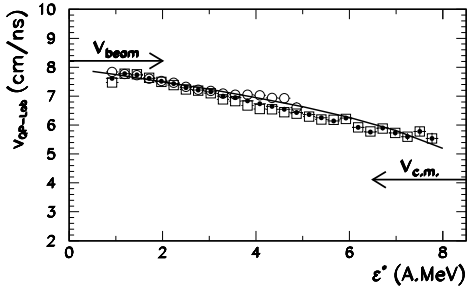}}
\caption{Velocity of the QP source as a function of the excitation
energy. Full and open points refer to Z$_1 >$ 50 and $<$ 50,
respectively. The dashed line is obtained by energy and momentum
conservation in a two body QP and QT kinematics~\cite{dag}.}
\label{twobody}
\end{center}
\end{figure}

In Fig.~\ref{twobody} we plot the velocity in the laboratory frame
of the QP source as a function of the excitation energy, with cuts
of Z$_1>$ 50 (full points) and Z$_1 <$ 50 (open points). The source
velocity expected for a two body (QP-QT) kinematics, obtained via
energy and momentum conservation in the hypothesis of an equal
sharing of the excitation energy by the two collision partners, is
shown by the dashed line~\cite{dag}. We can see that for all
calorimetrically reconstructed excitation energies, and for both
Z$_1$ cuts, the observed behavior is compatible with purely binary
collisions. This comparison shows that also the lighter Z$_1$'s,
corresponding to constant size QP remnants, do not come from central
reactions.

This discussion implies that we can safely consider the data as
characteristic of the de-excitation of constant size source in a
wide range of excitation energies.

Let us now come to the central question of data sorting. The global
distributions of Fig.~\ref{whole} reflect the excitation energy
deposit imposed by the dynamics of the entrance channel, and cannot
be considered as belonging to a single statistical ensemble. If a
sorting cannot be avoided, it is also clear that the shape of the
distributions will depend on the sorting choice.

The two de-excitation modes visible in Fig.~\ref{whole} are
associated to very different excitation energies. If they represent
two different phases, this means that the associated phase
transition should have a non zero latent heat, as it is the case for
regular liquid-gas. Therefore, the sorting variable should not
impose a too strong constraint on the deposited energy, such that
the two phases can be accessed in the same bin. In particular, in
the liquid-gas phase transition, Z$_1$ is known to be bimodal in the
canonical ensemble which allows huge energy fluctuations, while no
bimodality is observed in the microcanonical ensemble with fixed
energy~\cite{PRE}.

To search for a possible bimodal behavior, we should then in
principle sort the data in temperature bins, i.e. in canonical
ensembles. This is not possible experimentally, but we can choose a
sorting variable as close as possible to a canonical temperature.
Moreover, as suggested by previous papers~\cite{indra,rivet}, the
sorting  observable should better not be auto-correlated with
fragments and light particles emitted by the QP source. To fulfill
these requirements, as in~\cite{indra}, the transverse energy
Et$_{12}$ = $\sum_Z E_Z sin^2 (\theta_Z)$ of the light particles
($Z\leq2$) emitted by the quasi-target source has been chosen, which
is only loosely correlated to the QP observables. The QT had been
much larger than the QP, this sorting could be considered as a
canonical one. This is not our case, but it has still the advantage
to allow for relatively large energy fluctuations, as it is needed
to explore two phases that could be separated by a non zero latent
heat.

\begin{figure}[htbp]
\begin{center}
%\vspace{3cm}
\resizebox{0.7\columnwidth}{!}{\includegraphics{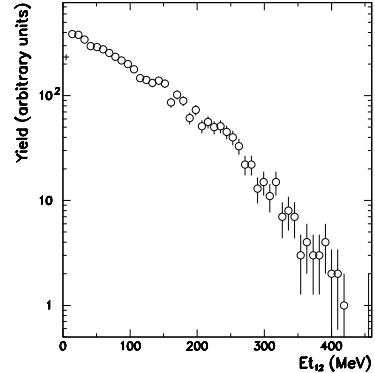}}
\caption{Transverse energy distribution of light charged particles
emitted in the backward hemisphere in the c. m. reference frame.}
\label{etra}
\end{center}
\end{figure}

It has to be noted that efficiency limitations for fragments in the
backward direction do not allow to have a total reconstruction of
the quasi-target source (see Fig.~\ref{vpar}). Therefore the
selection on QT characteristics can only be performed on light
particles; in particular the transverse energy Et$_{12}$ of these
particles is detected with good efficiency. The distribution of
Et$_{12}$ for the selected events is shown in Fig.~\ref{etra}.

This sorting can be assimilated to an impact parameter sorting. The
width of the transverse energy bins, equally spaced, was chosen of
40 MeV. Only the first six bins have sufficient statistics to be
considered for the subsequent analysis.

\begin{figure}[htbp]
\begin{center}
\resizebox{0.9\columnwidth}{!}{\includegraphics{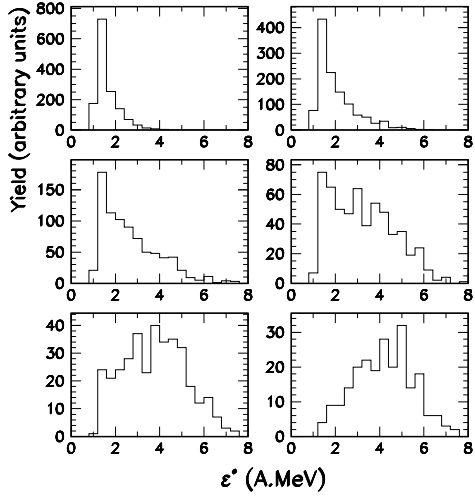}}
\caption{Calorimetric $\varepsilon$* distribution in the six bins of
transverse energy Et$_{12}$. Et$_{12}$ increases going from left to
right and from top to bottom panels.} \label{estar}
\end{center}
\end{figure}

The excitation energy constraint implied by this sorting is explored
in Fig.~\ref{estar}, which shows the distribution of the
calorimetric $\varepsilon$* in the six transverse energy bins. We
can notice from this figure that the variables Et$_{12}$ and
$\varepsilon$* are loosely correlated, and a relatively wide
distribution of $\varepsilon$* is obtained in most of the bins of
transverse energy. It is well known~\cite{reliability} that the
calorimetric measurement is not perfect, and the incomplete
detection creates a spurious width in the energy distribution. Since
this spurious width never exceed 1 MeV per nucleon, it is  clear
from Fig.\ref{estar} that the sorting in Et$_{12}$ bins cannot be
considered as a microcanonical selection, where no bimodality would
be expected.

\begin{figure}[htbp]
\begin{center}
\resizebox{0.9\columnwidth}{!}{\includegraphics{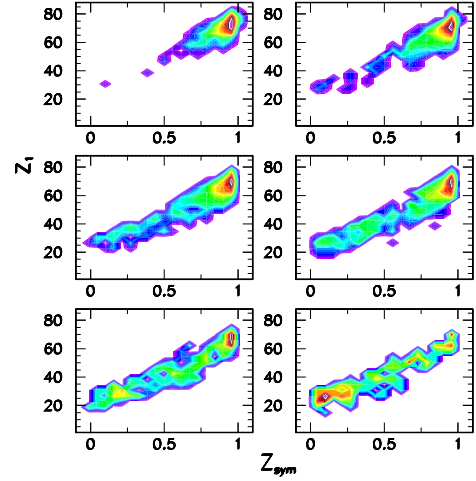}}
\caption{Charge of the heaviest fragment {\it vs.} the asymmetry
Z$_{sym}$ of the two heaviest in different bins of transverse
energy.} \label{bim}
\end{center}
\end{figure}

The charge of the heaviest fragment Z$_1$ is represented as a
function of the asymmetry Z$_{sym}$ of the two heaviest in
Fig.~\ref{bim}. We can see that the maximal probability does not
monotonically change with the centrality selection. The most
probable fragmentation pattern, characterized by a residue
exhausting most of the available charge and an important asymmetry
between the two largest fragments, abruptly changes between the
fourth and the fifth bin with the apparition of a second peak. This
peak represents multifragmentation events, with the largest fragment
comparable in size to the other emitted clusters. In the sixth bin
this second peak tends to become more prominent, even if the
situation would be more clear with higher statistics. Such a
discontinuous behavior agrees with the expectations from a phase
transition, and with the findings of the INDRA collaboration on
peripheral Xe + Sn and Au + Au collisions~\cite{indra}.

\begin{figure}[htbp]
\begin{center}
\resizebox{0.9\columnwidth}{!}{\includegraphics{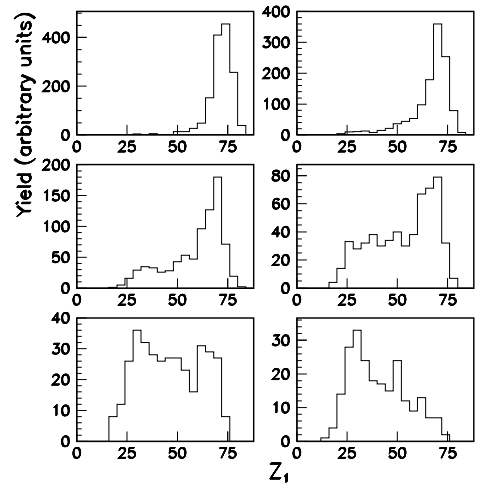}}
\caption{Distribution of the Z$_1$ variable in the different
transverse energy bins.} \label{projbig}
\end{center}
\end{figure}

\begin{figure}[htbp]
\begin{center}
\resizebox{0.9\columnwidth}{!}{\includegraphics{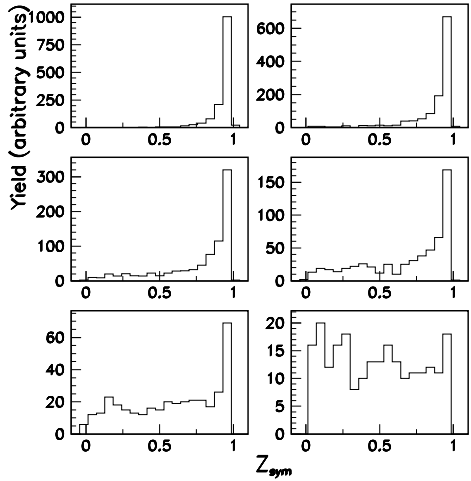}}
\caption{Distribution of the asymmetry Z$_{sym}$ in the different
transverse energy bins.} \label{projasym}
\end{center}
\end{figure}

Projecting the plots of Fig.~\ref{bim} onto the two axes, we can
note that the best indication of bimodality appears on the Z$_1$
variable (see Fig.~\ref{projbig}), whereas the plot of the asymmetry
Z$_{sym}$ does not show a clear bimodal behavior (see
Fig.~\ref{projasym}). The largest fragment size distribution, peaked
around $Z_1\approx$ 70 up to the fourth bin, shows a maximum around
$Z_1\approx 30$ in the sixth bin, passing through a configuration
(fifth bin) where a minimum in the probability appears to be
associated to the intermediate patterns, even if the statistics
should definitely be improved. This strongly suggests a first order
phase transition~\cite{PRE}.

\begin{figure}[htbp]
\begin{center}
\resizebox{0.99\columnwidth}{!}{\includegraphics{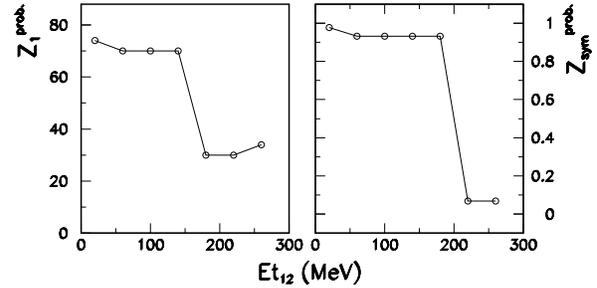}}
\caption{Evolution with Et$_{12}$ of the most probable value of
Z$_1$ (left panel) and  Z$_{sym}$ (right panel). The lines are drawn
to guide the eye. In this figure also the seventh bin of the
transverse energy is shown, despite of the low statistics.}
\label{jump}
\end{center}
\end{figure}

The indication of bimodality is less clear in the Z$_{sym}$
variable, which shows a wide distribution in the last bins without
any pronounced minimum ( see Fig.~\ref{projasym}). However the
sudden change in the most probable fragmentation pattern, shown in
the bidimensional correlations of Fig.~\ref{bim}, is clearly seen in
both variables. The most probable value of Z$_1$ and Z$_{sym}$ is
shown in Fig~\ref{jump} as a function of the transverse energy. Both
variables show a sudden decrease from the evaporation dominated
pattern, to the multifragmentation dominated one. This behavior is
again in agreement with the findings of ref.~\cite{indra} and
consistent with the expectation from a phase transition. In
principle a first order phase transition should be associated to a
discontinuous jump~\cite{PRE}, while a continuous power law behavior
would characterize a second order phase transition. This however
would be true only if the sorting variable could be assimilated to a
thermodynamical temperature. In the microcanonical ensemble, even a
first order transition is associated to a continuous behavior of the
order parameter. As a consequence, the power law behavior of the
average size of the largest cluster as a function of excitation
energy (see Fig.~\ref{cneg}), can be observed both in the case of a
critical behavior and in the coexistence zone of a first order phase
transition~\cite{big,richert}. In the experimental case, the
arbitrariness of the sorting and the absence of a physical external
bath does not allow to draw definite conclusions.

\begin{figure}[hb]
\begin{center}
\resizebox{0.99\columnwidth}{!}{\includegraphics{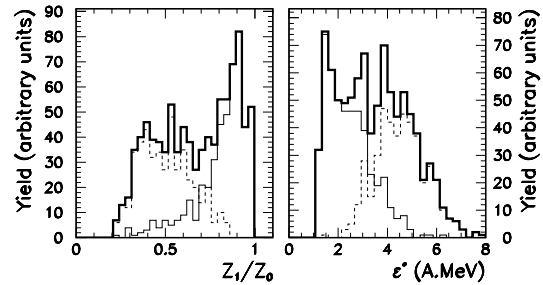}}
\caption{Distribution of Z$_1$/Z$_0$ (left) and $E^*$ (right) in the
fourth + fifth bins of transverse energy. Thick lines: global
distributions. Thin lines: $E^*<3.5 $A.MeV (left); $Z_1>50$ (right).
Dashed lines: $E^*>3.5$ A.MeV(left); $Z_1<50$ (right).}
\label{phase}
\end{center}
\end{figure}

A better understanding on the nature and order of the observed phase
change can be achieved form Fig.~\ref{phase}, which shows the
distribution of the largest cluster charge and excitation energy in
the Et$_{12}$ region (fourth + fifth bins) where the sudden change
in the fragmentation pattern is observed. As we have already
mentioned, a first order phase transition should imply a non zero
latent heat, meaning that the two "phases" observed at the same
"temperature" should be associated to different excitation energies.

We can see in Fig.~\ref{phase} that indeed the two decay modes
observed in the $Z_1$ distribution correspond to different values of
the calorimetric excitation energy: the cut $Z_1$=50 that roughly
identifies the two modes (see Fig.~\ref{whole}) succeeds in
splitting the energy distributions into two separate components
(right part of Fig.~\ref{phase}), even if the distance of the
centroids is too small to produce a clear bimodality in the
excitation energy distribution. The lower (higher) $Z_1$ component
roughly corresponds to a deposited energy higher (lower) than $3.5$
A.MeV (left part of Fig.~\ref{phase}). This behavior is consistent
with the expectations from a first order phase transition smoothed
by finite size effects.

If we interpret the two $Z_1$ distributions as two coexisting
phases, it would be tempting to estimate the latent heat of the
transition from the energy distance between the two peaks. The
"liquid" peak points to an excitation energy $E_1^*\approx 2$ A.MeV
which nicely agrees with the global $Z_1$-$E^*$ correlation shown in
the left part of Fig.~\ref{cneg}, and with the indication of the
fluctuation measurement shown in the right part of the same figure.
The "vapor" contribution peaks at $E_2^*\approx 5$ A.MeV, a lower
value respect to the location of the second divergence in the
fluctuation analysis. This discrepancy may be due to the intrinsic
limitations of the transverse energy sorting, that does not allow
sufficient energy fluctuations respect to a physical heat bath; it
may also point to an incomplete exploration of the high energy phase
space in our data sample, that cuts the distributions on the high
energy side.

To summarize, the results of Fig.\ref{phase} tend to suggest that
the observed sudden change from evaporation to multifragmentation
can be associated to first order phase transition. Higher statistics
samples obtained with collisions at higher beam energy could allow
to be conclusive about the compatibility between fluctuations and
bimodality~\cite{bonnet}. In addition a detailed study of the
convexity properties of the distributions is
needed~\cite{gul,bonnet2}.

\section{Conclusions}
\label{sec:3}

In this paper we have presented a new analysis of the 35 A.MeV
quasi-projectile Au+Au data collected with the Multics-Miniball
apparatus. The distributions of the largest cluster charge and of
the charge asymmetry between the two largest clusters detected in
each event have been studied. These data allow to analyze, with a
limited statistics, the de-excitation of a constant size
quasi-projectile source within a large range of dissipated energy. A
clear transition from an evaporative to a multifragmentation pattern
has been observed. The shape of the distributions have been studied
to search for a possible bimodal behavior, which would allow to
interpret this transition in the de-excitation mode as the finite
system counterpart of a first order phase transition. The asymmetry
distribution does not present a clear structure, while the largest
fragment charge appears bimodal.

The same data have shown several different signals that coherently
point to a first order liquid-gas-like phase transition. We recall
here the determination of thermodynamically consistent critical
exponents, both in a moment analysis~\cite{dag} and in an analysis
"\`a la Fisher"~\cite{critical}, and the fluctuation peak in the
partial energy distribution, with an absolute value of fluctuations
consistent with the existence of a negative branch for the
microcanonical heat capacity~\cite{reliability,negative}.

Concerning bimodality, the search of a convexity in the distribution
of the largest fragment emitted in each collisional event appears a
very direct, and therefore interesting signature of a first order
phase transition. The intrinsic weakness associated to this signal,
namely the arbitrariness in the choice of the sorting variable and
of the sorting bin width, will be the object of future
investigations~\cite{gul}. To confirm the significance of these
findings, it will be important to verify that the dynamics of the
entrance channel and the different bias associated to different
detectors do not influence the final results. To this aim, the
bimodality distributions for systems produced with different
entrance channels and detected with different apparatuses should be
compared in details. Moreover, the stability of the signals as a
function of the different ways of sorting the events should be
checked.

In particular, the qualitative trend of the data appears consistent
with the recent findings of the INDRA collaboration~\cite{indra} at
higher incident energies, but the behavior of the asymmetry variable
Z$_{sym}$ is not equivalent in the two data sets and deserves a
further analysis.

%
% BibTeX users please use
% \bibliographystyle{}
% \bibliography{}

\end{document}